\documentclass[12pt]{article}

\usepackage{graphicx,amsmath,amsfonts,amssymb,dcolumn,amsthm,slashed,bbm,xspace,pgffor,tikz,mathbbol,latexsym}
\usepackage[colorlinks,citecolor=blue,urlcolor=blue,linkcolor=blue]{hyperref}
\usepackage[utf8]{inputenc}
\usepackage[english]{babel}
\usepackage{lmodern}
\usepackage{microtype}
\usepackage[hang]{footmisc} 

\numberwithin{equation}{section}

\begin{document}

\title{\textbf{More about the renormalization properties of topological Yang-Mills theories}}

\author{\textbf{O.~C.~Junqueira$^1$}\thanks{octavio@if.uff.br}\ , \textbf{A.~D.~Pereira$^2$}\thanks{a.pereira@thphys.uni-heidelberg.de}\ , \textbf{G.~Sadovski$^{1}$}\thanks{gsadovski@id.uff.br}\  , \\ \textbf{R.~F.~Sobreiro$^1$}\thanks{rodrigo\_sobreiro@id.uff.br}\ , \textbf{A.~A.~Tomaz$^{3,4}$}\thanks{tomaz@cbpf.br}\\\\
\textit{{\small $^1$UFF - Universidade Federal Fluminense, Instituto de F\'isica,}}\\
\textit{{\small Av. Litorânea s/n, 24210-346, Niter\'oi, RJ, Brasil.}}\\
\textit{{\small $^2$Institute for Theoretical Physics, University of Heidelberg,}}\\
\textit{{\small Philosophenweg 12, 69120 Heidelberg, Germany}}\\
\textit{{\small $^3$CBPF - Centro Brasileiro de Pesquisas F\'isicas,}}\\
\textit{{\small Rua Dr. Xavier Sigaud 150, 22290-180, Rio de Janeiro, RJ, Brasil.}}\\
\textit{{\small $^4$Centro de Matemática, Computação e Cognição}}\\
\textit{{\small UFABC - Universidade Federal do ABC,}}\\
\textit{{\small 09210-580, Santo André, Brasil.}}}

\date{}
\maketitle
 
\begin{abstract}
Quantum properties of topological Yang-Mills theory in (anti-)self-dual Landau gauge were recently investigated by the authors. We extend the analysis of renormalizability for two generalized classes of gauges; each of them depending on one gauge parameter. The (anti-)self-dual Landau gauge is recovered at the vanishing of each gauge parameter. The theory shows itself to be renormalizable in these classes of gauges. Moreover, we discuss the ambiguity on the choice of the renormalization factors (which is not present in usual Yang-Mills theories) and argue a possible origin of such ambiguity.
\end{abstract}

\newpage
\tableofcontents
\newpage

\section{Introduction}

Topological quantum field theories are field theories that can be used to compute topological invariants. In four dimensions, the most celebrated example is perhaps the Donaldson-Witten (DW) theory \cite{Donaldson:1983xx,Donaldson:1987xx,Witten:1988ze,Birmingham:1991ty}. In \cite{Baulieu:1988xs}, Baulieu and Singer argued that DW theory can be casted in a BRST invariant apparel by defining a generalized set of BRST transformations which involve the Faddeev-Popov ghost field. In the Baulieu-Singer (BS) approach the DW action is obtained (along with extra Faddeev-Popov couplings and the gauge fixing for the gauge field) from the gauge fixing of the gauge-degrees-of-freedoms of a topologically invariant action. In this view, the gauge fixing choices open up the possibility of obtaining many different actions.

A particular appealing gauge choice in the BS approach is the (anti-)-self-dual Landau gauge ((A)SDLG), first introduced in \cite{Brandhuber:1994uf}. This choice encodes a rich set of Ward identities which allow to prove the renormalizability of the theory to all orders in perturbation theory\footnote{The proof is performed with the help of the algebraic renormalization technique \cite{Piguet:1995er}, which is independent of the renormalization scheme and valid to all orders in perturbation theory.} \cite{Brandhuber:1994uf,Junqueira:2017zea}. Moreover, the (A)SDLG has two crucial properties \cite{Junqueira:2017zea,Junqueira:2018xgl}: all Green functions are tree-level exact and; the gauge propagator vanishes identically.

In this work we generalize the (A)SDLG by introducing two gauge parameters. The modification relies in altering the Landau gauge condition on the gauge field to the linear covariant gauges and the (anti-)self-duality condition of the field strength to a non-(anti-)self-dual one, see \eqref{gf1} below. The gauge condition for the topological ghost remains the Landau transverse condition. It turns out that this gauge is not generally renormalizable. Nevertheless, we show that if we consider the linear covariant gauges and the non-(anti-)self-dual gauge separately, these gauges are indeed renormalizable to all orders in perturbation theory. It is important to be noted that, in both classes of gauges, the (A)SDLG is recovered by setting the gauge parameters to zero. The demonstration employs the algebraic renormalization technique \cite{Piguet:1995er}. It is worth mentioning that the vector supersymmetry \cite{Brandhuber:1994uf,Junqueira:2017zea} is not present in these new classes of gauges.

Beyond the renormalizability proof, we discuss the fact that the renormalization factors (the $Z$ factors) display a kind of freedom in their solution. It seems to be that, in these classes of gauges, there is a universal property allowing two free $Z$ factors. The origin of this freedom is also discussed and linked to the triviality of the BRST cohomology of topological Yang-Mills theories. Moreover, we use the gauge propagator as an example to show how some of the $Z$ factors are irrelevant in the renormalization of such objects.

The paper is organized as follows. In Section 2 we discuss the generalization of the (A)SDLG. Section 3 is devoted to the renormalizability properties of these gauges. In Section 4 we study the freedom on the choice of some $Z$ factors. Finally, Section 5   contains our concluding remarks and perspectives.

\section{Topological gauge theories in a generalized class of gauges}

Following \cite{Baulieu:1988xs}, a topological action $S_o[A]$ constructed with a $G$-valued gauge field $A_\mu^a$, where $G$ is a semi-simple Lie group, over an Euclidean flat four-dimensional spacetime is invariant under three infinitesimal gauge transformations, which can be written as
\begin{eqnarray}
\delta A^a_\mu &=& D_\mu^{ab}\alpha^b+\alpha_\mu^a\;, \label{gt1}\\
\delta F^a_{\mu\nu} &=& -gf^{abc}\alpha^bF_{\mu\nu}^c+D_\mu^{ab}\alpha_\nu^b-D_\nu^{ab}\alpha_\mu^b\;, \label{gt2}\\
\delta\alpha_\mu^a &=& D_\mu^{ab}\lambda^b\;,\label{gt3}
\end{eqnarray}
with $D^{ab}_\mu \equiv \delta^{ab}\partial_\mu - gf^{abc}A^{c}_{\mu}$ being the covariant derivative in the adjoint representation of the gauge group. The coupling parameter is $g$ while $\alpha^a$, $\alpha_\mu^a$ and $\lambda^a$ are the gauge parameters. The first transformation in equation \eqref{gt1} contains the usual gauge transformation parametrized by $\alpha^a$. The parameter $\alpha_\mu^a$ is $G$-valued and characterizes all other possible transformations associated to the fact that $S_o[A]$ is a topological invariant. Transformation \eqref{gt2} is a direct consequence of transformation \eqref{gt1}. Due to the presence of the covariant derivative terms in this transformation, the field strength, $F_{\mu\nu}^a=\partial_\mu A^a_\nu-\partial_\nu A^a_\mu+gf^{abc}A^b_\mu A^c_\nu$, acquires a gauge field character. Also from transformations \eqref{gt1}, it is easy to see that the gauge parameter $\alpha_\mu^a$ carries an extra ambiguity, characterized by the gauge transformation \eqref{gt3}.

The quantization of a topological gauge theory can be done by the BRST quantization method \cite{Baulieu:1988xs,Piguet:1995er}. The first step is to promote the gauge parameters to ghost fields: $\alpha^a\longrightarrow c^a$, the Faddeev-Popov ghost; $\alpha_\mu^a\longrightarrow\psi^a_\mu$, the topological ghost and; $\lambda^a\longrightarrow\phi^a$, the ghost of the ghost (or bosonic ghost). The corresponding BRST transformations can be defined as
\begin{eqnarray}
sA_\mu^a&=&-D_\mu^{ab}c^b+\psi^a_\mu\;,\nonumber\\
sc^a&=&\frac{g}{2}f^{abc}c^bc^c+\phi^a\;,\nonumber\\
s\psi_\mu^a&=&gf^{abc}c^b\psi^c_\mu+D_\mu^{ab}\phi^b\;,\nonumber\\
s\phi^a&=&gf^{abc}c^b\phi^c\;,\label{brst1}
\end{eqnarray}
where $s$ is the nilpotent BRST operator. The second step is to fix the gauge freedom with suitable gauge constraints: for the gauge field we employ the linear covariant gauge condition; for the field strength, a non-(anti-)self-dual gauge condition is chosen and; for the topological ghost we set the Landau gauge constraint; namely\footnote{It should be noted that $\alpha$ and $\beta$ must be necessarily negative quantities. Otherwise, the Boltzmann factor would be with the wrong sign in the path integral.},
\begin{eqnarray}
\partial_\mu A_\mu^a&=&-\alpha b^a\;,\nonumber\\
F^a_{\mu\nu}\pm\widetilde{F}_{\mu\nu}^a&=&-\beta B^a_{\mu\nu}~,\nonumber\\
\partial_\mu\psi^a_\mu&=&0\;,\label{gf1}
\end{eqnarray}
where $\widetilde{F}_{\mu\nu}^a$ is the dual field strength defined as $\widetilde{F}_{\mu\nu}^a=\frac{1}{2}\epsilon_{\mu\nu\alpha\beta}F_{\alpha\beta}^a$ and $\alpha$ and $\beta$ are gauge parameters. To enforce such constraints, we introduce three BRST doublets:
\begin{eqnarray}
s\bar{c}^a&=&b^a\;,\;\;\;\;\;\;\;\;sb^a\;=\;0\;,\nonumber\\
s\bar{\chi}^a_{\mu\nu}&=&B_{\mu\nu}^a\;,\;\;\;sB_{\mu\nu}^a\;=\;0\;,\nonumber\\
s\bar{\phi}^a&=&\bar{\eta}^a\;,\;\;\;\;\;\;\;s\bar{\eta}^a\;=\;0\;,\label{brst2}
\end{eqnarray}
where $\bar{\chi}^a_{\mu\nu}$ and $B_{\mu\nu}^a$ are (anti-)self-dual fields, according to the the positive (negative) sign in the second condition in \eqref{gf1}. In \eqref{brst2}, the fields $b^a$, $B^a_{\mu\nu}$ and $\bar{\eta}^a$ are Lautrup-Nakanishi fields which implement the gauge conditions \eqref{gf1} while $\bar{c}^a$, $\bar{\chi}^a_{\mu\nu}$ and $\bar{\phi}^a$ are the Faddeev-Popov anti-ghost, the topological anti-ghost and the bosonic anti-ghost fields, respectively. For completeness and further use, the quantum numbers of all fields are displayed in Table \ref{table1}. The complete gauge fixing action in the gauge choices \eqref{gf1} takes the form
\begin{eqnarray}
S_{gf}&=&s\int d^4z\left[\bar{c}^a\left(\partial_\mu A_\mu^a+\frac{\alpha}{2} b^a\right)+\frac{1}{2}\bar{\chi}^a_{\mu\nu}\left(F_{\mu\nu}^a\pm\widetilde{F}_{\mu\nu}^a + \frac{\beta}{2}B^a_{\mu\nu}\right)+\bar{\phi}^a\partial_\mu\psi^a_\mu\right]\nonumber\\
&=&\int d^4z\left[b^a(\partial_\mu A_\mu^a+\frac{\alpha}{2}b^a)+\frac{1}{2}B^a_{\mu\nu}\left(F_{\mu\nu}^a\pm\widetilde{F}_{\mu\nu}^a + \frac{\beta}{2}B^a_{\mu\nu}\right)+
\left(\bar{\eta}^a-\bar{c}^a\right)\partial_\mu\psi^a_\mu+\bar{c}^a\partial_\mu D_\mu^{ab}c^b+\right.\nonumber\\
&-&\left.\frac{1}{2}gf^{abc}\bar{\chi}^a_{\mu\nu}c^b\left(F_{\mu\nu}^c\pm\widetilde{F}_{\mu\nu}^c\right)-\bar{\chi}^a_{\mu\nu}\left(\delta_{\mu\alpha}\delta_{\nu\beta}\pm\frac{1}{2}\epsilon_{\mu\nu\alpha\beta}\right)D_\alpha^{ab}\psi_\beta^b+\bar{\phi}^a\partial_\mu D_\mu^{ab}\phi^b+\right.\nonumber\\
&+&\left.gf^{abc}\bar{\phi}^a\partial_\mu\left(c^b\psi^c_\mu\right)\right]\;.\label{gfaction}
\end{eqnarray}

\begin{table}[h]
\centering
\setlength{\extrarowheight}{.5ex}
\begin{tabular}{cc@{\hspace{-.3em}}cccccccccc}
\cline{1-1} \cline{3-12}
\multicolumn{1}{|c|}{Field} & & \multicolumn{1}{|c|}{$A$} & \multicolumn{1}{c|}{$\psi$} & \multicolumn{1}{c|}{$c$} & \multicolumn{1}{c|}{$\phi$} & \multicolumn{1}{c|}{$\bar{c}$} & \multicolumn{1}{c|}{$b$} &\multicolumn{1}{c|}{$\bar{\phi}$} & \multicolumn{1}{c|}{$\bar{\eta}$} & \multicolumn{1}{c|}{$\bar{\chi}$} & \multicolumn{1}{c|}{$B$}   
\\ \cline{1-1} \cline{3-12} 
\\[-1.18em]
\cline{1-1} \cline{3-12}
\multicolumn{1}{|c|}{Dim} & & \multicolumn{1}{|c|}{1} & \multicolumn{1}{c|}{1} & \multicolumn{1}{c|}{0} & \multicolumn{1}{c|}{0} & \multicolumn{1}{c|}{2} & \multicolumn{1}{c|}{2} &\multicolumn{1}{c|}{2} & \multicolumn{1}{c|}{2} & \multicolumn{1}{c|}{2} & \multicolumn{1}{c|}{2}   
\\
\multicolumn{1}{|c|}{Ghost n$^o$} & & \multicolumn{1}{|c|}{0} & \multicolumn{1}{c|}{1} & \multicolumn{1}{c|}{1} & \multicolumn{1}{c|}{2} & \multicolumn{1}{c|}{-1} & \multicolumn{1}{c|}{0} &\multicolumn{1}{c|}{-2} & \multicolumn{1}{c|}{-1} & \multicolumn{1}{c|}{-1} & \multicolumn{1}{c|}{0}   
\\ \cline{1-1} \cline{3-12} 
\end{tabular}
\caption{Quantum numbers of the fields.}
\label{table1}
\end{table}

By integrating out the auxiliary field $B^a_{\mu\nu}$, a Yang-Mills term is produced. It is worth noting that this is not a genuine Yang-Mills term because it is multiplied by a gauge parameter and is originated from a BRST variation, \emph{i.e.}, it belongs to the trivial sector of the cohomology of $s$ \cite{Baulieu:1988xs,Piguet:1995er}. The tree-level gauge propagator is easily computed, 
\begin{equation}\label{TGP}
\langle A^{a}_\mu A^{b}_\nu\rangle_0(p) = \delta^{ab}\left[\frac{\beta}{4p^2}\left(\delta_{\mu\nu}-\frac{p_{\mu}p_{\nu}}{p^2}\right)+\frac{\alpha}{p^2}\frac{p_\mu p_\nu}{p^2}\right]\;,
\end{equation}
which is a pure gauge propagator. This is consistent with the fact that topological gauge theories carry only global physical degrees of freedom. In fact, the unphysical nature of the gauge field as a local object is even more appealing at the (A)SDLG, where $\alpha=\beta=0$ and the gauge propagator \eqref{TGP} vanishes \cite{Brandhuber:1994uf,Junqueira:2017zea}. This property, being a very peculiar result for this gauge choice, has a strong consequence: all connected $n$-point Green functions are tree-level exact \cite{Junqueira:2018xgl}.

\section{Renormalizability}

\subsection{Generalities}\label{GENs}

To study the renormalizability of the action \eqref{gfaction}, the first step is to introduce some external sources in order to control the non-linear nature of the BRST transformations and other possible non-linear symmetries. The required BRST doublets are \cite{Junqueira:2017zea}
\begin{eqnarray}
s\tau^a_\mu&=&\Omega_\mu^a\;,\;\;\;\;\;\;\;\;s\Omega_\mu^a\;=\;0\;,\nonumber\\
sE^a&=&L^a\;,\;\;\;\;\;\;\;\;\;sL^a\;=\;0\;,\nonumber\\
s\Lambda_{\mu\nu}^a&=&K_{\mu\nu}^a\;,\;\;\;\;sK_{\mu\nu}^a\;=\;0\;,\label{brst3}
\end{eqnarray}
with the corresponding quantum numbers displayed in the Table \ref{table2} below. The external action accounting for all relevant non-linear operators is 
\begin{eqnarray}
S_{ext}&=&s\int d^4z\left(\tau_\mu^aD_\mu^{ab}c^b+\frac{g}{2}f^{abc}E^ac^bc^c+gf^{abc}\Lambda^a_{\mu\nu}c^b\bar{\chi}^c_{\mu\nu}\right)\nonumber\\
&=&\int d^4z\left[\Omega_\mu^aD_\mu^{ab}c^b+\frac{g}{2}f^{abc}L^ac^bc^c+gf^{abc}K^a_{\mu\nu}c^b\bar{\chi}^c_{\mu\nu}+\tau^a_\mu\left(D_\mu^{ab}\phi^b+gf^{abc}c^b\psi_\mu^c\right)+\right.\nonumber\\
&+&\left.gf^{abc}E^ac^b\phi^c+gf^{abc}\Lambda^a_{\mu\nu}c^bB^c_{\mu\nu}-gf^{abc}\Lambda^a_{\mu\nu}\phi^b\bar{\chi}^c_{\mu\nu}-\frac{g^2}{2}f^{abc}f^{bde}\Lambda^a_{\mu\nu}\bar{\chi}^c_{\mu\nu}c^dc^e\right]\;.\label{extaction}
\end{eqnarray}
Hence, the full action to be considered is 
\begin{equation}
\Sigma=S_o[A]+S_{gf}+S_{ext}\;.\label{fullaction}
\end{equation}

\begin{table}[h]
\centering
\setlength{\extrarowheight}{.5ex}
\begin{tabular}{cc@{\hspace{-.3em}}cccccc}
\cline{1-1} \cline{3-8}
\multicolumn{1}{|c|}{Source} & & \multicolumn{1}{|c|}{$\tau$} & \multicolumn{1}{c|}{$\Omega$} & \multicolumn{1}{c|}{$E$} & \multicolumn{1}{c|}{$L$} & \multicolumn{1}{c|}{$\Lambda$} & \multicolumn{1}{c|}{$K$}
\\ \cline{1-1} \cline{3-8} 
\\[-1.18em]
\cline{1-1} \cline{3-8}
\multicolumn{1}{|c|}{Dim} & & \multicolumn{1}{|c|}{3} & \multicolumn{1}{c|}{3} & \multicolumn{1}{c|}{4} & \multicolumn{1}{c|}{4} & \multicolumn{1}{c|}{2} & \multicolumn{1}{c|}{2} \\
\multicolumn{1}{|c|}{Ghost n$^o$} & & \multicolumn{1}{|c|}{-2} & \multicolumn{1}{c|}{-1} & \multicolumn{1}{c|}{-3} & \multicolumn{1}{c|}{-2} & \multicolumn{1}{c|}{-1} & \multicolumn{1}{c|}{0}
\\ \cline{1-1} \cline{3-8}
\end{tabular}
\caption{Quantum numbers of the external sources.}
\label{table2}
\end{table}

The analysis of renormalizability of the action \eqref{fullaction} can be performed within the algebraic renormalization framework \cite{Piguet:1995er}. Specifically, one adds to the classical action \eqref{fullaction} a counterterm $\Sigma^c$,
\begin{equation}
\Gamma^{(1)}=\Sigma+\epsilon\Sigma^c\;,\label{qation1}
\end{equation}
where $\epsilon$ is a small parameter, given by the most general functional which is polynomial in the fields and their derivatives, sources and parameters of dimension 4 and vanishing ghost number. Hence, one imposes to $\Gamma^{(1)}$ all Ward identities obeyed by $\Sigma$. The action \eqref{fullaction} is said to be renormalizable if the surviving counterterm can be absorbed by the classical action by means of multiplicative redefinition of the fields, sources and parameters as
\begin{equation}
\Sigma(\Phi_0,J_0,\xi_0)=\Sigma(\Phi,J,\xi)+\epsilon\Sigma^c(\Phi,J,\xi)\;,\label{abs1}
\end{equation}
where
\begin{eqnarray}
\Phi_0&=&Z^{1/2}_\Phi\Phi\;,\;\;\;\Phi\in\{\mathrm{all\;fields}\}\;,\nonumber\\
J_0&=&Z_JJ\;,\;\;\;J\in\{\mathrm{all\;sources}\}\;,\nonumber\\
\xi_0&=&Z_\xi\xi\;,\;\;\;\xi\in\{g,\alpha,\beta\}\;.\label{abs2}
\end{eqnarray}

The action \eqref{fullaction} is not renormalizable in general. A long but straightforward computation leads to a counterterm that can not be absorbed by the classical action \eqref{fullaction}. For instance, a term of the form $gf^{abc}\frac{\delta\Sigma}{\delta\phi^a}c^bc^c$ appears in $\Sigma^c$. Nevertheless, there are three special cases of \eqref{fullaction} in which all divergences can be absorbed: the case $\alpha=\beta=0$ which is the (A)SDLG; the case $\alpha\ne0$ and $\beta=0$, called $\alpha$-gauges and; the case $\alpha=0$ and $\beta\ne0$, called $\beta$-gauges. Let us start our discussion with the special case of the (A)SDLG ($\alpha=\beta=0$), whose quantum properties were previously studied in \cite{Junqueira:2017zea,Junqueira:2018xgl}.

\subsection{(Anti-)self-dual Landau gauges}

In the case of the (A)SDLG, the action \eqref{fullaction} reduces to $\Sigma_{(A)SDLG}=\Sigma(\alpha=\beta=0)$. This is the most symmetric case due to the rich set of Ward identities displayed by this gauge, see \cite{Junqueira:2017zea,Junqueira:2018xgl}. In particular, the symmetry \eqref{Tfull1} and the vector supersymmetry \cite{Brandhuber:1994uf,Junqueira:2017zea} play a fundamental role. The renormalizability of this gauge was established in \cite{Brandhuber:1994uf,Junqueira:2017zea} and the most general counterterm reads
\begin{eqnarray}
\Sigma^{c}_{(A)SDLG} &=& S_\Sigma \int d^4x \, a \, \bar{\chi}^a_{\mu\nu} F^a_{\mu\nu}\;,\nonumber\\
&=&\int d^4x \;a\left(B^a_{\mu\nu} F^a_{\mu\nu} - 2\bar{\chi}^a_{\mu\nu}D^{ab}_\mu\psi^b_\nu - gf^{abc}\bar{\chi}_{\mu\nu}^ac^bF^c_{\mu\nu}\right)\;,\label{GCT}
\end{eqnarray}
where $a$ is an independent renormalization coefficient. This counterterm can indeed be reabsorbed in the classical action $\Sigma(\alpha=\beta=0)$ by means of multiplicative redefinition of the fields, sources and parameters according to \eqref{abs1} and \eqref{abs2}. The resulting $Z$ factors obey the following system of equations:
\begin{eqnarray}
Z^{1/2}_A &=&Z_b^{-1/2}=Z_g^{-1}\;,\nonumber\\ Z^{1/2}_{\bar{c}}&=&Z^{1/2}_{\bar{\eta}}= Z_\psi^{-1/2} = Z_\Omega = Z^{-1/2}_c \;,\nonumber\\
Z_{\bar{\phi}}^{1/2}&=&Z_\phi^{-1/2}=Z_\tau=Z_L=Z^{-1}_gZ^{-1}_c\;,\nonumber\\ Z_E&=&Z_g^{-2}Z^{-3/2}_c\;,\nonumber\\
Z_K&=&Z_g^{-1}Z_c^{-1/2}Z_{\bar{\chi}}^{-1/2}\;,\nonumber\\
Z_\Lambda &=&Z_g^{-2}Z_c^{-1}Z_{\bar{\chi}}^{-1/2}\;,\nonumber\\
Z^{1/2}_B Z^{1/2}_A &=& Z^{1/2}_{\bar{\chi}} Z^{1/2}_c  =1 + \epsilon a\;.\label{Z1a}
\end{eqnarray}
As observed in \cite{Junqueira:2017zea}, this system is self-consistent. However, it is clearly undetermined because there are fifteen equations and seventeen fields, sources and parameters. It means there are two free $Z$ factors, characterizing an ambiguity in the renormalization of the theory. For instance, the system \eqref{Z1a} in the way we have written it, can be completely fixed by suitably choosing $Z_g$ and $Z_c$. We will return to this issue later in Section \ref{ZFAC}.

\subsection{$\alpha$-gauges}

Now, let us consider the case where $\beta = 0$ while keeping $\alpha$ arbitrary in the action \eqref{fullaction}, the $\alpha$-gauges. The full action is now
\begin{equation}\label{sigma3}
\Sigma_\alpha=\Sigma|_{\beta=0}\;.
\end{equation}
The proof of renormalizability is established in the Appendix \ref{ALFA}. It turns out that the most general counterterm is also given by \eqref{GCT}. The $\alpha$-gauges also show themselves to be stable by means of \eqref{Z1a} supplemented by the renormalization factor of the gauge parameter $\alpha$,
\begin{equation}
Z_\alpha^{1/2}=Z_g^{-1}\;.\label{Z1}
\end{equation}
We gain one more equation for the system of equation determining the $Z$ factors but we also gain an extra $Z$ factor, $Z_\alpha$. Hence, the ambiguity remains.

\subsection{$\beta$-gauges}\label{betagauges}

The third case we study is characterized by setting $\alpha=0$ and maintaining an arbitrary $\beta$ in the original action \eqref{fullaction}, the $\beta$-gauges. The full action is then
\begin{equation}\label{sigma4}
\Sigma_\beta=\Sigma\vert_{\alpha=0}~.
\end{equation}
This action is also renormalizable, as discussed in Appendix \ref{BETA}, and the most general counterterm assumes the form \eqref{GTC3}. An interesting feature to be observed at this point (which also occurs at the (A)SDLG and the $\alpha$-gauges) is that the Faddeev-Popov term does not appear in the counterterm \eqref{GTC3}. This implies that $Z_gZ_A^{1/2}=1$. Using this information in the terms $a_1B\partial A$ and $a_2gBAA$ of \eqref{GTC3}, one finds $a_2=a_1/2$. Then, the counterterm \eqref{GTC3} is simplified to
\begin{eqnarray}
\Sigma^c_\beta &=& S_\Sigma \int d^4x \left(a\, \bar{\chi}^a_{\mu\nu} F_{\mu\nu}^a+ \tilde{a}\beta  \bar{\chi}^a_{\mu\nu}B^a_{\mu\nu}\right)\nonumber\\
&=& \int d^4x \left[a\left(B^a_{\mu\nu} F^a_{\mu\nu} - 2\bar{\chi}^a_{\mu\nu}D^{ab}_\mu\psi^b_\nu - gf^{abc}\bar{\chi}_{\mu\nu}^ac^bF^c_{\mu\nu}\right) + \frac{\tilde{a}}{2}\beta B^a_{\mu\nu} B^a_{\mu\nu}
\right]~,\label{GTC3a}
\end{eqnarray}
where we have renamed the renormalization constants as $a=a_1/2$ and $\tilde{a}/2=a_4$. All relations between the $Z$ factors can be straightforwardly found from \eqref{abs1}, \eqref{abs2} and \eqref{GTC3a}. The result preserves the system formed by \eqref{Z1a} and \eqref{Z1} with the additional equation
\begin{equation}\label{Z3a}
Z_\beta Z_B = 1 + \epsilon\tilde{a}\;.
\end{equation}

Again, an extra equation is gained together with an extra $Z$ factor, $Z_\beta$. For this reason the ambiguity persists.

\section{Discussing the $Z$ factors}\label{ZFAC}

In the previous section, we have discussed the algebraic renormalization properties in three classes of gauges, namely, the (A)SDLG and the $\alpha$- and $\beta$-gauges, respectively. In particular, the (A)SDLG can be obtained from the latter classes by continuous deformations, \emph{i.e.}, $\alpha\rightarrow0$ or $\beta\rightarrow0$. In all cases the action is renormalizable to all orders in perturbation theory. However, the system of $Z$ factors is, in all cases, undetermined. The number of equations $n$ and the number of variables $z$ (the $Z$ factors) are related by $z=n+2$ in all three cases. It seems that there is a kind of freedom in the choice of two of the $Z$ factors. We will now discuss this ambiguity in more details.

\subsection{$Z$ factors in each gauge}

Let us analyze the (A)SDLG first. As commented before, this gauge has two peculiar features \cite{Junqueira:2018xgl}: 1) all $n$-point Green functions are tree-level exact and; 2) the gauge propagator vanishes to all orders in perturbation theory. This means that there are no radiative corrections to be implemented. In this case, it is evident that the ambiguity in the $Z$ factors \eqref{Z1a} is irrelevant: it does not matter how we choose the $Z$ factors since there are no renormalizations to be made. Essentially, all divergent diagrams are canceled out by the vanishing of the gauge propagator \cite{Junqueira:2018xgl}. In principle, we can choose $Z_g$ and $Z_c$ as we wish. For instance, we can take the simplest choice
\begin{equation}
Z_g=Z_c=1\;,\label{Zb1}
\end{equation}
leading to the fact that all $Z$ factors are equal to 1 except for
\begin{equation}
Z_B^{1/2}=Z_{\bar{\chi}}^{1/2}=Z_\Lambda^{-1}=Z_K^{-1}=1+\epsilon a\;.\label{Zb2}
\end{equation}
This choice is particularly appealing because it implies that the $\beta$-function and the gauge propagator anomalous dimension vanish. 

If we focus now on the $\alpha$-gauges, we observe that the system of equations determining the $Z$ factors are the same as the system of the (A)SDLG with the extra equation \eqref{Z1} determining $Z_\alpha$. It is easy to see that the proposal \eqref{Zb1} is also consistent in this case, providing, again, \eqref{Zb2} and $Z_\alpha=1$. Thus, the vanishing of the $\beta$-function remains valid for the $\alpha$-gauges. 

Turning to the $\beta$-gauges, the system of equations \eqref{Z1a} are enriched with an extra equation \eqref{Z3a} determining $Z_\beta$. Again, the choice \eqref{Zb1} is consistent, leading to \eqref{Zb2} and
\begin{equation}
Z_\beta=1+\epsilon(\tilde{a}-2a)\;.\label{Zb3}
\end{equation}
It seems that the vanishing of the $\beta$-function and the behavior of the anomalous dimensions are the same in the analyzed gauges if we set \eqref{Zb1}.

\subsection{Comparison with Yang-Mills theories}

To understand more closely the origin of such ambiguities, we must observe that the set of symmetries in the gauges analyzed eliminates the kinetic term of the Faddeev-Popov ghost at the counterterms. Because of this, we get
\begin{equation} \label{Zc}
Z_c Z_{\bar{c}}=1\;.
\end{equation}
From the gauge-ghost vertex ($\bar{c}Ac$), which is also absent in the counterterm\footnote{In Yang-Mills theories quantized at the Landau gauge this property is known as the \emph{non-renormalization of the gluon-ghost vertex} \cite{Blasi:1990xz,Piguet:1995er}. The same result is obtained here for a more general class of gauges.}, and the relation \eqref{Zc}, we achieve 
\begin{equation}\label{ZgA}
Z_g Z^{1/2}_A =1\;.
\end{equation}
The two relations \eqref{Zc} and \eqref{ZgA} are decoupled, in other words, only by determining $Z_c$ or $Z_{\bar{c}}$ we do not get any information about $Z_g$ or $Z_A$. Nevertheless, the factor $Z_A$ could be individually determined if the classical action had a kinetic term for the gauge field. In the usual Yang-Mills theory, where the term $F^a_{\mu\nu}F^a_{\mu\nu}$ is present, $Z_A$ can be directly determined  from the gauge field kinetic term. But in topological Yang-Mills theories there are no kinetic terms for the gauge field. By this fact, the determination of $Z_A$ becomes impossible.

The same analysis we did for the Faddeev-Popov ghost terms can be performed for the bosonic ghost term, leading to
\begin{equation}
Z_{\bar{\phi}}Z_\phi=1\;.\label{Zphi}
\end{equation}
From the $\bar{\phi}A\phi$ vertex we also obtain \eqref{ZgA}.

For any other interacting term including $A$, $g$ also appears, making the combination $gA$ or $g^2A^2$ to be irrelevant due to \eqref{ZgA}. Moreover, the mixed propagators encoding $A$ also do not give any extra information. The analysis for the source terms also does not help (these terms always include an extra variable for each new relation between $Z$s.). Ultimately, one can infer that \eqref{Zc} and \eqref{ZgA} are the main basic relations that could solve the puzzle. Essentially, we need two extra informations about the $Z$-factors which are not encoded in the system \eqref{Z1a}. It is not difficult to conclude that the absence of a Yang-Mills term in the original action is the origin of the ambiguity of the $Z_A$ factor.

Another feature in the usual Yang-Mills theories (quantized in the Landau gauge) is that $Z_c = Z_{\bar{c}}$ which relies on the discrete symmetry
\begin{eqnarray}\label{cbarc}
c^a &\longrightarrow & \bar{c}^a\;,\nonumber \\
\bar{c}^a &\longrightarrow & -{c}^a\;.
\end{eqnarray}
This condition, together with the Faddeev-Popov ghost kinetic term, are sufficient to determine $Z_c$ and $Z_{\bar{c}}$. It is easy to see that the action \eqref{fullaction} does not obey such symmetry\footnote{It is instructive to observe that discrete symmetries between the other ghosts of topological Yang-Mills theories ($\phi^a$ and $\bar{\phi}^a$ and; $\psi^a_\mu$ and $\bar{\chi}^a_{\mu\nu}$) are also not present in \eqref{fullaction}.}, which explains the second ambiguity.

In essence we can infer that the difference between YM theories and TYM theories relies in their  cohomology properties. The non-trivial character of YM cohomology enables extra equations to determine $Z_A$ and $Z_c$. Moreover, a non-trivial cohomology implies on local physical degrees of freedom whose renormalization affect physical observables. Thus, a freedom in the choice of some renormalization factors could affect physical observables in catastrophic ways. On the other hand, the trivial nature of TYM cohomology is associated with the fact that all local degrees of freedom are non-physical (see \eqref{TGP} for instance - the gauge field propagator is totally gauge dependent) and such kind of freedom in how some objects renormalize can be interpreted as a reflex of the cohomology triviality.

\subsection{Gauge field propagators}

The ambiguity can also be understood by looking at the gauge field propagators. For instance, at the (A)SDLG, the gauge field propagator vanishes to all orders in perturbation theory \cite{Junqueira:2018xgl}. We immediately find that we have a liberty to choose any $Z_A$ we want: take $\langle A^a_\mu A^b_\nu \rangle_R$ as the dressed propagator and $\langle A^a_\mu A^b_\nu \rangle_0$ the bare one. Thus, 
\begin{equation}
\langle A^a_\mu(x) A^b_\nu(y) \rangle_R = Z_A \langle A^a_\mu(x) A^b_\nu(y) \rangle_0 = 0 \quad \Rightarrow \quad \langle A^a_\mu(x) A^b_\nu(y) \rangle_0 = 0,
\end{equation}
independently of $Z_A$. 

In the $\alpha$-gauges we found that $Z_\alpha=Z_A$, see \eqref{Z1a} and \eqref{Z1}. The expression of the tree-level gluon propagator at the $\alpha$-gauges is easily computed, 
\begin{equation} \label{ZAZA}
\langle A^a_\mu A^b_\nu \rangle_0(p) = \delta^{ab} \frac{\alpha}{p^2} \frac{p_\mu p_\nu}{p^2}\;.
\end{equation}
Therefore, after the redefinitions of the fields and parameters and using \eqref{Z1a} and \eqref{Z1}, $Z_A$ is canceled at both sides of \eqref{ZAZA}. Again, we conclude that we have the liberty to choose any renormalization factor for the gauge field. 

The $\beta$-gauges is no different from the previous cases. From \eqref{Z1a} and \eqref{Z3a} one obtains
\begin{eqnarray}
Z_\beta = Z_A\left[1+2\epsilon \left(\tilde{a}-2a\right)\right]\;.\label{Zrel2}
\end{eqnarray}
Now, the tree-level gluon propagator takes the form
\begin{equation}
\langle A^{a}_\mu A^{b}_\nu\rangle_0(p) = \delta^{ab}\frac{\beta}{4p^2}\left(\delta_{\mu\nu}-\frac{p_{\mu}p_{\nu}}{p^2}\right)\,.\label{ZAZA2}
\end{equation}
Once again, after the renormalizations, the factor $Z_A$ is canceled at both sides of \eqref{ZAZA2}. 

From the gauge field propagator, the freedom in the choice of $Z_A$ is clearly illustrated. As a consequence of the first equation in \eqref{Z1a}, \emph{i.e.} $Z_A=Z_g^{-1}$, this freedom is transmitted to the renormalization of the coupling parameter.

\section{Conclusions}

We studied the renormalizability of topological Yang-Mills theories for generalized classes of gauges which encodes the (A)SDLG. Differently from the (A)SDLG, these classes have the particular property that the gauge field propagator is non-vanishing. These generalizations were performed by the introduction of  two gauge parameters ($\alpha$ and $\beta$) which allowed us to write a general linear covariant gauge function for the gauge field and a non-(anti-)self-dual gauge condition for the field strength, as described in equations \eqref{gf1}.

The first case, characterized by non-vanishing $\alpha$ and $\beta$ - see \eqref{fullaction} -, is not renormalizable since the counterterm carries some interactions which are not present in the classical action. However, the theory is renormalizable for the special cases where one of the gauge parameters vanishes (for $\beta=0$, we defined the $\alpha$-gauges and for $\alpha=0$ we worked within the $\beta$-gauges). We recall that the (A)SDLG is recovered for both gauge classes by setting $\alpha$ or $\beta$ to zero. The proof of renormalizability was performed employing the algebraic renormalization techniques \cite{Piguet:1995er}, which is valid to all orders in perturbation theory.

In the case of the $\alpha$-gauges we have found that the counterterm is the same as in the (A)SDLG which carries only one independent renormalization parameter. Thus, the renormalization factors are determined by the same set of self-consistent equations as the (A)SDLG, namely \eqref{Z1a} supplemented with \eqref{Z1}, which determines the renormalization of $\alpha$.

For the $\beta$-gauges, an extra independent renormalization appears and the counterterm is a bit more general than the one for the $\alpha$-gauges, see \eqref{GTC3a}. It turns out that the renormalization factors system is also given by \eqref{Z1a}, with the extra equation determining the renormalization of $\beta$ given by \eqref{Z3a}. 

An interesting feature of the (A)SDLG is that, due to the exact vanishing of the gauge propagator, radiative corrections are completely absent \cite{Junqueira:2018xgl}. In the $\alpha$- and $\beta$-gauges, however, this property might not be true since the gauge propagator does not vanish, see \eqref{TGP}. In fact, even if by some surprising result the gauge propagator is tree-level exact, it could contribute to loop diagrams and the tree-level exactness of the (A)SDLG would be lost in these generalized gauges.

Besides the renormalization proof of the generalized gauges, we found that, in all cases (including the (A)SDLG) the number of equations $n$ determining the $Z$ factors and the number of renormalization factors $z$ always relate through $n=z+2$. Hence, there is a kind of freedom in two $Z$ factors. In the way we wrote equations \eqref{Z1a}, all $Z$ factors are determined by the choice of $Z_g$ and $Z_c$. An example of interesting solution is $Z_g=Z_c=1$, leading to a vanishing $\beta$-function. In this case, the divergences are absorbed by $\bar{\chi}$, $B$ and $\beta$.

In order to illustrate the freedom in the $Z$ factors, we analyzed the gauge propagator in each case ((A)SDL, $\alpha$- and, $\beta$-gauges). We have shown that the gauge propagator renormalization is independent from $Z_A$, which means that the choice of the gauge field renormalization factor does not affect the quantum properties of the gauge propagator.

Finally, we have provided a possible interpretation of this renormalization freedom in terms of the BRST cohomology and by comparison with the usual Yang-Mills theories. Essentially, since the BRST of topological Yang-Mills theories define a trivial cohomology, there are no local physical degrees of freedom, as should be expected from a topological field theory. Thus, the way we choose to renormalize these degrees of freedom seems to be irrelevant for the global degrees of freedom. Moreover, by comparison with ordinary Yang-Mills theories, the absence of a Yang-Mills term in the non-trivial sector of the cohomology and, the absence of kinetic terms for the ghosts in the counterterms we found, appear to rely on the origin of the renormalization freedom of the theory.

\section*{Acknowledgements}

The Conselho Nacional de Desenvolvimento Cient\'ifico e Tecnol\'ogico (CNPq - Brazil) and the Coordena\c{c}\~ao de Aperfei\c{c}oamento de Pessoal de N\'ivel Superior (CAPES) are acknowledged for financial support.

\appendix

\section{Renormalizability proof of the $\alpha$-gauges}\label{ALFA}

The aim of this first appendix is to prove the renormalizability of the action \eqref{sigma3}, \emph{i.e.} the renormalizability of the topological Yang-Mills theories at the $\alpha$-gauges. With this purpose, we will employ the algebraic renormalization technique \cite{Piguet:1995er}.

The action \eqref{sigma3} displays a few Ward identities:
\begin{itemize}
\item Slavnov-Taylor identity due the BRST invariance:
\begin{equation}
\mathcal{S}(\Sigma_\alpha)=0\;,\label{st1}
\end{equation}
where
\begin{eqnarray} \label{ST}
\mathcal{S}(\Sigma_\alpha)&=&\int d^4z\left[\left(\psi^a_\mu-\frac{\delta\Sigma_\alpha}{\delta\Omega^a_\mu}\right)\frac{\delta\Sigma_\alpha}{\delta A^a_\mu}+\frac{\delta\Sigma_\alpha}{\delta\tau^a_\mu}\frac{\delta\Sigma_\alpha}{\delta\psi^a_\mu}+\left(\phi^a+\frac{\delta\Sigma_\alpha}{\delta L^a}\right)\frac{\delta\Sigma_\alpha}{\delta c^a}+\frac{\delta\Sigma_\alpha}{\delta E^a}\frac{\delta\Sigma_\alpha}{\delta\phi^a}+\right.\nonumber\\
&+&\left.b^a\frac{\delta\Sigma_\alpha}{\delta\bar{c}^a}+\bar{\eta}^a\frac{\delta\Sigma_\alpha}{\delta\bar{\phi}^a}+B^a_{\mu\nu}\frac{\delta\Sigma_\alpha}{\delta\bar{\chi}^a_{\mu\nu}}+\Omega^a_\mu\frac{\delta\Sigma_\alpha}{\delta\tau^a_\mu}+L^a\frac{\delta\Sigma_\alpha}{\delta E^a}+K^a_{\mu\nu}\frac{\delta\Sigma_\alpha}{\delta\Lambda^a_{\mu\nu}}\right]\;.\label{st2}
\end{eqnarray}

\item Gauge fixing and anti-ghost equations:
\begin{equation} \label{gf}
\frac{\delta \Sigma_\alpha}{\delta b^a} = \partial_\mu A^a_\mu + \alpha b^a\;; \quad \quad \frac{\delta \Sigma_\alpha}{\delta \bar{c}^a} - \partial_\mu\frac{\delta \Sigma_\alpha}{\delta \Omega^a_\mu}= -\partial_\mu \psi^a_\mu\;.
\end{equation} 

\item Second gauge fixing and anti-ghost equations:
\begin{equation}\label{2gf}
\frac{\delta \Sigma_\alpha}{\delta \bar{\eta}^a} = \partial_\mu \psi^a_\mu\;; \quad \quad \frac{\delta \Sigma_\alpha}{\delta \bar{\phi}^a} - \partial_\mu \frac{\delta \Sigma_\alpha}{\delta \tau^a_\mu} = 0\;.
\end{equation}

\item First non-linear bosonic symmetry:
\begin{equation}
T^{(1)}(\Sigma_\alpha) = 0\;, \label{T1}
\end{equation}
where
\begin{eqnarray}\label{T1a}
T^{(1)}(\Sigma_\alpha) &=& \int d^4z \left[ \frac{\delta \Sigma_\alpha}{\delta \Omega^a_\mu} \frac{\delta\Sigma_\alpha}{\delta \psi^a_\mu} + \left(\phi^a - \frac{\delta\Sigma_\alpha}{\delta L^a}\right)\frac{\delta\Sigma_\alpha}{\delta \phi^a} + c^a\frac{\delta\Sigma_\alpha}{\delta c^a} - \bar{\phi}^a\frac{\delta\Sigma_\alpha}{\delta\bar{\phi}^a} - \bar{\eta}^a\left(\frac{\delta\Sigma_\alpha}{\delta \bar{\eta}^a}+ \frac{\delta\Sigma_\alpha}{\delta \bar{c}^a}\right)+\right.\nonumber\\
&-&\left. \Omega^a_\mu \frac{\delta\Sigma_\alpha}{\Omega^a_\mu} - \tau^a_\mu \frac{\delta\Sigma_\alpha}{\tau^a_\mu} - 2L^a \frac{\delta\Sigma_\alpha}{\delta L^a} - 2E^a \frac{\delta\Sigma_\alpha}{\delta E^a} -K^a_{\mu\nu}\frac{\delta\Sigma_\alpha}{\delta K^a_{\mu\nu}} - \Lambda^a_{\mu\nu} \frac{\delta\Sigma_\alpha}{\delta \Lambda^a_{\mu\nu}} \right]\;.
\end{eqnarray}

\item Second non-linear bosonic symmetry:
\begin{equation}
T^{(2)}\left(\Sigma_\alpha\right)=0\;,\label{T2a}
\end{equation}
where
\begin{eqnarray}\label{T2b}
T^{(2)}(\Sigma_\alpha)&=&\int d^4z \left[ \frac{\delta \Sigma_\alpha}{\delta K^a_{\mu\nu}} \frac{\delta\Sigma_\alpha}{\delta B^a_{\mu\nu}} + c^a \frac{\delta\Sigma_\alpha}{\delta c^a} - \bar{c}^a\left(\frac{\delta\Sigma_\alpha}{\delta \bar{c}^a} + \frac{\delta\Sigma_\alpha}{\delta \bar{\eta}^a}\right) + \phi^a\frac{\delta\Sigma_\alpha}{\delta \phi^a}-\bar{\phi}^a\frac{\delta\Sigma_\alpha}{\delta \bar{\phi}^a} +\right.\nonumber\\
&-& \left.\Omega^a_\mu \frac{\delta\Sigma_\alpha}{\delta\Omega^a_\mu} - \tau^a_\mu \frac{\delta\Sigma_\alpha}{\delta\tau^a_\mu} - 2L^a \frac{\delta\Sigma_\alpha}{\delta L^a} - 2E^a \frac{\delta\Sigma_\alpha}{\delta E^a} - \Lambda^a_{\mu\nu} \frac{\delta\Sigma_\alpha}{\delta\Lambda^a_{\mu\nu}} - K^a_{\mu\nu} \frac{\delta\Sigma_\alpha}{\delta K^a_{\mu\nu}} \right]\;.
\end{eqnarray}

\item Global ghost supersymmetry:
\begin{equation}
\mathcal{G}_3\Sigma_\alpha=0\;,\label{gss1}
\end{equation}
where
\begin{equation}\label{G3}
\mathcal{G}_3=\int d^4z\left[\bar{\phi}^a\left(\frac{\delta}{\delta\bar{\eta}^a}+\frac{\delta}{\delta\bar{c}^a}\right)-c^a\frac{\delta}{\delta\phi^a}+\tau^a_\mu\frac{\delta}{\delta\Omega^a_\mu}+2E^a\frac{\delta}{\delta L^a} + \Lambda^a_{\mu\nu}\frac{\delta}{\delta K^a_{\mu\nu}}\right].
\end{equation}

\end{itemize}

We notice that, just like the (A)SDLG, the symmetries $T^{(1)}$ and $T^{(2)}$, in \eqref{T1} and \eqref{T2a}, respectively, can be combined to compose a more suitable symmetry operator,
\begin{eqnarray}
T(\Sigma_\alpha)=T^{(1)}(\Sigma_\alpha)-T^{(2)}(\Sigma_\alpha)=0\;,\label{Tfull1}
\end{eqnarray}
such that
\begin{equation}
T(\Sigma_\alpha)=\int d^4z\left[\frac{\delta\Sigma_\alpha}{\delta\Omega^a_\mu}\frac{\delta\Sigma_\alpha}{\delta\psi^a_\mu}-\frac{\delta\Sigma_\alpha}{\delta L^a}\frac{\delta\Sigma_\alpha}{\delta\phi^a}-\frac{\delta\Sigma_\alpha}{\delta K^a_{\mu\nu}}\frac{\delta\Sigma_\alpha}{\delta B^a_{\mu\nu}}+\left(\bar{c}^a-\bar{\eta}^a\right)\left(\frac{\delta\Sigma_\alpha}{\delta\bar{c}^a}+\frac{\delta\Sigma_\alpha}{\delta\bar{\eta}^a}\right)\right]\;.
\label{Tfull2}
\end{equation}

To study the perturbative quantum stability of action \eqref{sigma3} one adds to the classical action \eqref{sigma3} the most general counterterm $\Sigma^c_\alpha$ by means of 
\begin{equation}
\Gamma^{(1)}=\Sigma_\alpha+\epsilon\Sigma^c_\alpha\;.\label{ct1}
\end{equation}
The counterterm $\Sigma^c_\alpha$ is an integrated polynomial in the fields, sources and their derivatives, and parameters with dimension bounded by 4 and vanishing ghost number. Following the algebraic renormalization technique \cite{Piguet:1995er}, we impose that all Ward identities valid for the classical action \eqref{sigma3} can be extended to the quantum action \eqref{ct1}. Hence, the counterterm $\Sigma^c_\alpha$ should satisfy the following constraints
\begin{eqnarray}
\mathcal{S}_{\Sigma_\alpha}\Sigma_\alpha^c&=&0\;,\label{c1}\\
\frac{\delta \Sigma^c_\alpha}{\delta b^a}&=&0\;,\label{c2}\\
\frac{\delta \Sigma^c_\alpha}{\delta \bar{c}^a} - \partial_\mu \frac{\delta \Sigma^c_\alpha}{\delta \Omega^a_\mu}&=&0\;,\label{c3}\\
\frac{\delta \Sigma^c_\alpha}{\delta \bar{\eta}^a}&=&0\;,\label{c4}\\
\frac{\delta \Sigma^c_\alpha}{\delta \bar{\phi}^a} - \partial_\mu \frac{\delta \Sigma^c_\alpha}{\delta \tau^a_\mu}&=&0\;,\label{c5}\\
T_{\Sigma_\alpha} \Sigma^c_\alpha&=&0\;,\label{c6}\\
\mathcal{G}_3\Sigma^c_\alpha&=& 0\;,\label{c7}
\end{eqnarray}
where the linearized Slavnov-Taylor operator is given by
\begin{eqnarray}
\mathcal{S}_{\Sigma_\alpha}&=&\int d^4z\left[\left(\psi^a_\mu-\frac{\delta\Sigma_\alpha}{\delta\Omega^a_\mu}\right)\frac{\delta}{\delta A^a_\mu}-\frac{\delta\Sigma_\alpha}{\delta A^a_\mu}\frac{\delta}{\delta\Omega^a_\mu}+\frac{\delta\Sigma_\alpha}{\delta\tau^a_\mu}\frac{\delta}{\delta\psi^a_\mu}+\left(\Omega^a_\mu+\frac{\delta\Sigma_\alpha}{\delta\psi^a_\mu}\right)\frac{\delta}{\delta\tau^a_\mu}+\right.\nonumber\\
&+&\left.\left(\phi^a+\frac{\delta\Sigma_\alpha}{\delta L^a}\right)\frac{\delta}{\delta c^a}+\frac{\delta\Sigma_\alpha}{\delta c^a}\frac{\delta}{\delta L^a}+\frac{\delta\Sigma_\alpha}{\delta E^a}\frac{\delta}{\delta\phi^a}+\left(L^a+\frac{\delta\Sigma_\alpha}{\delta\phi^a}\right)\frac{\delta}{\delta E^a}+\right.\nonumber\\
&+&\left. b^a\frac{\delta}{\delta\bar{c}^a}+\bar{\eta}^a\frac{\delta}{\delta\bar{\phi}^a}+B^a_{\mu\nu}\frac{\delta}{\delta\bar{\chi}^a_{\mu\nu}}+K^a_{\mu\nu}\frac{\delta}{\delta\Lambda^a_{\mu\nu}}\right]\;,\label{st3}
\end{eqnarray}
and the linearized bosonic symmetry operator is
\begin{eqnarray}
T_{\Sigma_\alpha}&=&\int d^4z \left[ \frac{\delta \Sigma_\alpha}{\delta \Omega^a_\mu} \frac{\delta}{\delta \psi^a_\mu}+\frac{\delta \Sigma_\alpha}{\delta \psi^a_\mu}\frac{\delta}{\delta\Omega^a_\mu}-\frac{\delta\Sigma_\alpha}{\delta L^a}\frac{\delta}{\delta \phi^a} - \frac{\delta \Sigma_\alpha}{\delta \phi^a}\frac{\delta}{\delta L^a}-\frac{\delta\Sigma_\alpha}{\delta K^a_{\mu\nu}}\frac{\delta}{\delta B^a_{\mu\nu}}-\frac{\delta\Sigma_\alpha}{\delta B^a_{\mu\nu}}\frac{\delta}{\delta K^a_{\mu\nu}}+\right.\nonumber\\
&+&\left.\left(\bar{c}^a- \bar{\eta}^a\right)\left(\frac{\delta}{\delta \bar{c}^a}+ \frac{\delta}{\delta \bar{\eta}^a}\right)\right]\;.
\end{eqnarray}
Since the operator $S_{\Sigma_\alpha}$ is nilpotent, it defines a cohomology problem for $\Sigma^c_\alpha$. The cohomology is trivial and the Slavnov-Taylor identity is free of anomalies \cite{Brandhuber:1994uf,Junqueira:2017zea}. Hence, the general solution of \eqref{c1} is
\begin{equation}
\Sigma^c=\mathcal{S}_{\Sigma_\alpha}\Delta^{(-1)}\;,\label{ct2}
\end{equation}
where $\Delta^{(-1)}$ is an integrated local polynomial in the fields, sources and their derivatives, and parameters bounded by dimension 4 and ghost number -1. From \eqref{ct2} and the constraints \eqref{c2} -- \eqref{c7}, it is straightforward to conclude that the most general counterterm allowed is given by \eqref{GCT}. To check if the the $\alpha$-gauges are stable is to check if the counterterm \eqref{GCT} can be reabsorbed by the classical action $\Sigma_\alpha$ by means of the redefinition of the fields, sources and parameters as in \eqref{abs1} and \eqref{abs2}. It is easy to check that the solution is given by \eqref{Z1a} and \eqref{Z1}, which completes the proof of renormalizability of topological Yang-Mills quantized at the $\alpha$-gauges.

\section{Renormalizability proof of the $\beta$-gauges}\label{BETA}

The renormalizability proof of topological Yang-Mills theories at the $\beta$-gauges follows the same procedures of the $\alpha$-gauges discussed in the previous appendix. The starting action is now \eqref{sigma4} and its symmetries are described by the following Ward identities:
\begin{itemize}
\item Slavnov-Taylor identity:
\begin{equation}
\mathcal{S}(\Sigma_\beta)=0\;,\label{st1B}
\end{equation}
where
\begin{equation} \label{STB}
\mathcal{S}(\Sigma_\beta)=\mathcal{S}(\Sigma_\alpha)\big|_{\Sigma_\alpha\rightarrow\Sigma_\beta}\;,
\end{equation}
where $\mathcal{S}(\Sigma_\alpha)$ was defined in \eqref{st1}.

\item Gauge fixing and anti-ghost equations:
\begin{equation} \label{gfB}
\frac{\delta \Sigma_\beta}{\delta b^a} = \partial_\mu A^a_\mu\;; \quad \quad \frac{\delta \Sigma_\beta}{\delta \bar{c}^a} - \partial_\mu\frac{\delta \Sigma_\beta}{\delta \Omega^a_\mu}= -\partial_\mu \psi^a_\mu\;.
\end{equation} 

\item Second gauge fixing and anti-ghost equations:
\begin{equation}\label{2gfB}
\frac{\delta \Sigma_\beta}{\delta \bar{\eta}^a} = \partial_\mu \psi^a_\mu\;; \quad \quad \frac{\delta \Sigma_\beta}{\delta \bar{\phi}^a} - \partial_\mu \frac{\delta \Sigma_\beta}{\delta \tau^a_\mu} = 0\;.
\end{equation}

\item First non-linear bosonic symmetry:
\begin{equation}
T^{(1)}(\Sigma_\beta) = 0\;, \label{T1B}
\end{equation}
where
\begin{equation}\label{T1aB}
T^{(1)}(\Sigma_\alpha)=T^{(1)}(\Sigma_\alpha)\big|_{\Sigma_\alpha\rightarrow\Sigma_\beta}\;,
\end{equation}
where $T^{(1)}(\Sigma_\alpha)$ was defined in \eqref{T1a}.

\item Bosonic ghost equation:
\begin{equation}\label{bgeB}
\mathcal{G}^a_\phi\Sigma_\beta= \Delta^a_\phi\;,
\end{equation}
where
\begin{eqnarray}
\mathcal{G}^a_\phi&=&\int d^4z\left(\frac{\delta}{\delta\phi^a}-gf^{abc}\bar{\phi}^b\frac{\delta}{\delta b^c}\right),\nonumber \\
\Delta_\phi^a&=&gf^{abc}\int d^4z\left(\tau_\mu^bA_\mu^c+E^bc^c+\Lambda_{\mu\nu}^b\bar{\chi}_{\mu\nu}^c\right)\;.
\label{bg2B}
\end{eqnarray}

\item Second Faddeev-Popov ghost equation:
\begin{equation}
\mathcal{G}^a_2\Sigma_\beta=\Delta^a\;,\label{Sg1B}
\end{equation}
where
\begin{eqnarray}
\mathcal{G}^a_2 &=& \int d^4z\left[\frac{\delta}{\delta c^a}-gf^{abc}\left(\bar{\phi}^b\frac{\delta}{\delta\bar{c}^c}+A^b_\mu\frac{\delta}{\delta\psi^c_\mu}+c^b\frac{\delta}{\delta\phi^c}-\bar{\eta}^b\frac{\delta}{\delta b^c}+E^b\frac{\delta}{\delta L^c} +\tau^b_\mu \frac{\delta}{\delta \Omega^c_\mu}\right)\right]\;,\nonumber\\
\Delta^a&=&gf^{abc}\int d^4z\left(E^b\phi^c-\Omega^b_\mu A_\mu^c-\tau^b_\mu\psi_\mu^c-L^bc^c+\Lambda^b_{\mu\nu}B^c_{\mu\nu}-K^b_{\mu\nu}\bar{\chi}^c_{\mu\nu}\right)\;.\label{Sg2B}
\end{eqnarray}

\item Global ghost supersymmetry:
\begin{equation}
\mathcal{G}_3\Sigma_\beta=0\;,\label{gss1B}
\end{equation}
where
\begin{equation}\label{G3B}
\mathcal{G}_3=\int d^4z\left[\bar{\phi}^a\left(\frac{\delta}{\delta\bar{\eta}^a}+\frac{\delta}{\delta\bar{c}^a}\right)-c^a\frac{\delta}{\delta\phi^a}+\tau^a_\mu\frac{\delta}{\delta\Omega^a_\mu}+2E^a\frac{\delta}{\delta L^a} + \Lambda^a_{\mu\nu}\frac{\delta}{\delta K^a_{\mu\nu}}\right].
\end{equation}

\end{itemize}

The perturbative quantum stability of action \eqref{sigma4} is studied just like the previous case. We start by adding to the classical action \eqref{sigma4} the most general counterterm $\Sigma^c_\beta$ by means of 
\begin{equation}
\Gamma^{(1)}=\Sigma_\beta+\epsilon\Sigma^c_\beta\;.\label{ct1B}
\end{equation}
The counterterm $\Sigma^c_\beta$ is, again, an integrated polynomial in the fields, sources and their derivatives, and parameters with dimension bounded by 4 and vanishing ghost number. Then, we impose the validity of all Ward identities valid for the classical action \eqref{sigma4} to the quantum action \eqref{ct1B}. Therefore, the counterterm $\Sigma^c_\beta$ must satisfy the following constraints
\begin{eqnarray}
\mathcal{S}_{\Sigma_\beta}\Sigma_\beta^c&=&0\;,\label{c1B}\\
\frac{\delta \Sigma^c_\beta}{\delta b^a}&=&0\;,\label{c2B}\\
\frac{\delta \Sigma^c_\beta}{\delta \bar{c}^a} - \partial_\mu \frac{\delta \Sigma^c_\beta}{\delta \Omega^a_\mu}&=&0\;,\label{c3B}\\
\frac{\delta \Sigma^c_\beta}{\delta \bar{\eta}^a}&=&0\;,\label{c4B}\\
\frac{\delta \Sigma^c_\beta}{\delta \bar{\phi}^a} - \partial_\mu \frac{\delta \Sigma^c_\beta}{\delta \tau^a_\mu}&=&0\;,\label{c5B}\\
T^{(1)}_{\Sigma_\beta} \Sigma^c_\beta&=&0\;,\label{c6B}\\
\mathcal{G}^a_\phi\Sigma^c_\beta&=& 0\;,\label{c7B}\\
\mathcal{G}^a_2\Sigma^c_\beta&=& 0\;,\label{c8B}\\
\mathcal{G}_3  \Sigma^c_\beta  &=& 0\;,\label{c9B}
\end{eqnarray}
where the linearized Slavnov-Taylor operator is given by
\begin{equation}
\mathcal{S}_{\Sigma_\beta}=\mathcal{S}_{\Sigma_\alpha}\big|_{\Sigma_\alpha\rightarrow\Sigma_\beta}\;,\label{LSTB}
\end{equation}
where $\mathcal{S}_{\Sigma_\alpha}$ was defined in \eqref{st3} and $T^{(1)}_{\Sigma_\beta}$ is given by 
\begin{eqnarray}\label{T1linB}
T^{(1)}_{\Sigma_\beta}&=&\int d^4x \left[ \frac{\delta \Sigma_\beta}{\delta \Omega^a_\mu} \frac{\delta}{\delta \psi^a_\mu} + \left(\phi^a - \frac{\delta\Sigma_\beta}{\delta L^a}\right)\frac{\delta}{\delta \phi^a} + c^a\frac{\delta}{\delta c^a} - \bar{\phi}^a\frac{\delta}{\delta\bar{\phi}^a} - \bar{\eta}^a\left(\frac{\delta}{\delta \bar{c}^a}+ \frac{\delta}{\delta \bar{\eta}^a}\right)+\right.\nonumber\\
&+&\left.  \left(\frac{\delta \Sigma_\beta}{\delta \psi^a_\mu}- \Omega^a_\mu\right) \frac{\delta}{\delta\Omega^a_\mu} - \tau^a_\mu \frac{\delta}{\tau^a_\mu} - \left(\frac{\delta \Sigma_\beta}{\delta \phi^a}+2L^a\right) \frac{\delta}{\delta L^a} - 2E^a \frac{\delta}{\delta E^a} -K^a_{\mu\nu}\frac{\delta}{\delta K^a_{\mu\nu}} +\right. \nonumber\\
&-&\left.\Lambda^a_{\mu\nu} \frac{\delta }{\delta \Lambda^a_{\mu\nu}}\right]\;.
\end{eqnarray}

The operator $S_{\Sigma_\beta}$ is also nilpotent. Henceforth, it defines a cohomology problem for $\Sigma^c_\beta$. The trivial BRST cohomology implies that the general solution of \eqref{c1B} is
\begin{equation}
\Sigma^c_\beta=\mathcal{S}_{\Sigma_\beta}\Delta^{(-1)}\;,\label{ct2B}
\end{equation}
where $\Delta^{(-1)}$ is an integrated local polynomial in the fields, sources and their derivatives, and parameters bounded by dimension 4 and ghost number -1. From \eqref{ct2B} and the constraints \eqref{c2B} -- \eqref{c9B}, it is straightforward to show that the most general counterterm allowed is actually given by
\begin{eqnarray}
\Sigma^c_\beta &=& S_\Sigma \int d^4x \left( a_1 \, \bar{\chi}^a_{\mu\nu} \partial_\mu A^a_{\nu} + a_2 \, g f^{abc} \bar{\chi}^a_{\mu\nu} A^b_\mu A^c_{\nu}+ a_{4}\beta \bar{\chi}^a_{\mu\nu}B^a_{\mu\nu}\right)\nonumber\\
&=& \int d^4x \left\{ a_1\left[B^a_{\mu\nu} \partial_\mu A^a_{\nu} - \bar{\chi}^a_{\mu\nu}\partial_\mu\left(\psi^a_\nu - \frac{\delta \Sigma}{\delta \Omega^a_{\nu}}\right)\right] + \right.\nonumber\\
&+&\left.a_2\left[gf^{abc} B^a_{\mu\nu} A^b_{\mu} A^c_{\nu} - 2g f^{abc}\bar{\chi}^a_{\mu\nu}\left(\psi^b_\mu - \frac{\delta \Sigma}{\delta \Omega^b_\mu}\right) A^c_\nu\right] + a_{4}\beta B^a_{\mu\nu} B^a_{\mu\nu}
\right\}~.\label{GTC3}
\end{eqnarray}
As discussed in Section \ref{betagauges}, the analysis of the quantum stability of the $\beta$-gauges via \eqref{abs1} and \eqref{abs2} leads to the relation $a_2=a_1/2$. This simplification reduces \eqref{GTC3} to \eqref{GTC3a}. The solution for the $Z$ factors are given by \eqref{Z1a} and \eqref{Z3a} and the proof of renormalizability of topological Yang-Mills quantized at the $\beta$-gauges is complete.

\bibliographystyle{utphys2}
\bibliography{library}

\end{document}